# Design of Gallium Nitride Resonant Cavity Light Emitting Diodes on Si Substrates**

By Michael A. Mastro, Joshua D. Caldwell, Ron T. Holm, Rich L. Henry, and Charles R. Eddy Jr.*

The efficiency of semiconductor light emitting diodes (LEDs) has continuously improved since the first fabrication of an infrared device in 1960. The key metric has been the external quantum efficiency, which describes the number of photons externally radiated per injected electron. Rapid initial progress was achieved in enhancing the carrier injection efficiency through bandgap engineering in heterostructure and, subsequently, quantum well devices.[1] The second significant evolution entailed the development of sophisticated crystal growth equipment and techniques to improve the crystal quality of the LED.[2] Defect levels in the III-arsenide material system have improved to the point that the fraction of recombination that is non-radiative now approaches zero.[1] Still, defect-assisted, non-radiative recombination continues to be problematic in the III-nitride material system, which is typically fabricated on non-native substrates.

The current evolution in LED design involves enhancing the external extraction of photons generated in the active region.[3] Compared to the scientific solutions presented above, improvements in extraction efficiency typically involve costly engineering solutions including substrate thinning or removal, flip-chip bonding, surface roughening, and die shaping. Additionally, a reflector cup is often employed to redirect laterally emitted light; however, this package decreases the brightness per area.[1] An elegant solution is the use of a resonant cavity (RCLED) to guide the preferential propagation of generated photons into the light extraction cone.[3] The compact and directed nature of a RCLED allows improvements in directionality, spectral purity, bandwidth, and overall extraction efficiency that are not available from a traditional LED.

This letter presents a novel theoretical and experimental design for a III-nitride RCLED grown on low-cost Si substrates. This development presents an opportunity to significantly improve the efficiency and cost structure of III-nitride based LEDs.

A RCLED can overcome the inherent limitation of total internal reflection associated with planar LEDs, where only light that is emitted into the escape cone can be extracted. Light propagating towards the surface at an angle greater than critical angle is subjected to total internal reflection and may not be extracted ($\theta_c = \sin^{-1}(n_{air}/n_s)$, where $n_s$ is the index of refraction of the semiconductor).[4] Thus, the external extraction efficiency of an isotropic planar LED is quite low with $\eta = 0.04$ for an index of $n_{GaN} = 2.5$.

An RCLED is one solution to redistribute the spontaneous emission pattern into the desired extraction cone. In such a structure, a light emitting region with an optical thickness on the order of the emission wavelength is surrounded by two mirrors. The bottom mirror consists of a high reflectance ($R_2 > 95\%$) multi-layer Bragg mirror or single-layer metal layer. The top mirror is of lower reflectance ($40\% > R_1 > 70\%$) to allow

---

[*] Dr. M.A. Mastro, Dr. J.D. Caldwell, Dr. R.T. Holm, Dr. R.L. Henry, Dr. C.R. Eddy Jr.

U.S. Naval Research Laboratory,

[**] Research at the Naval Research Lab is partially supported by the Office of Naval Research.

greater transmission. Thus, a multi-layer Bragg mirror is typically used for the top mirror to minimize absorption, although a very thin multi-layer metallic mirror can also be used as will be described in this letter. This structure creates a Fabry-Perot cavity with an optical length of

$$\lambda_{FP} = \frac{2n_s L_c}{m_c}, \quad [1]$$

where $L_c$ is the cavity thickness and $m_c$ is the cavity order.[1] The roundtrip phase change of the mode is[5]

$$\Phi(\theta) = \frac{4\pi n_s L_c}{m_c}\cos\theta, \quad [2]$$

which is dependent on the angle θ from the normal to the surface.[6] The mode extracted through the light cone is maximized at an angle $\theta_o$, which is obtained from the Airy function of the roundtrip phase change,[7-9]

$$\Phi(\theta_0) = 2\pi m_c. \quad [3]$$

Thus, the extraction efficiency is inversely proportional to the mode order, $\eta_{ext} \approx 1/m_c$. This efficiency is further decreased by the broadening of the Airy function by the spectral spread, σ, of the source ($\lambda_s$),[9]

$$\Delta\Phi = \Phi(\theta)\frac{\Delta\lambda}{\lambda_s} = \Phi(\theta)\left[\frac{\sigma}{\lambda_s} + \frac{1}{Fm_c}\right]. \quad [4]$$

The inverse finesse term, 1/F, of the cavity is typically insignificant for a weak-coupling planar micro-cavity[9] as is discussed in this article. For comparison, strong-coupling can be found in the 3-dimensional cavity created in low-dimensional optical nanostructures.[10] To maximize extraction, the mode of the cavity is detuned to place the maximum emission angle between the surface normal and the critical angle for total internal reflection.[8]

Proper cavity design creates interference effects in the Fabry-Perot resonator that alters the internal angular power distribution. The propagation direction of the photons is forced into the extraction cone and away from the total internal reflection regime. Following Benisty et al.,[1,6-8] the electromagnetic wave propagating within a multilayer Fabry-Perot cavity can be decomposed into TE and TM components. Accordingly, the emitted intensity with the internal emission angle (θ) based on the horizontal or vertical orientation as well as the s or p polarization of the source plane-wave component, $A^{or,pol}$, is given by

$$I_{dip}^{or,pol}(\theta) = \frac{\left|A_\uparrow^{or,pol} + A_\downarrow^{or,pol} r_2 e^{-i2\phi_{2eff}}\right|^2}{\left|1 - |r_1 r_2| e^{-i2\phi_{eff}}\right|^2}, \quad [5]$$

where $r_1$ and $r_2$ are the top and bottom reflection coefficients.[7] The total phase shift, $\phi=\phi_1+\phi_2$, and local phase shift above ($\phi_1$) and below ($\phi_2$) the emission dipole are related to the local distance from the dipole to the mirror ($d_1,d_2$) by $\phi_i = k_0 n_s d_i \cos\theta$, where $k_0$ is the free space wavevector. The local distance must include the effective penetration depth of the mirror, particularly for a distributed Bragg reflector (DBR) structure, where the reflectance is, in effect, a weighted distributed sum of the reflectance from each interface.[11]

The total distribution of the emission pattern is decoupled as

$$I(\theta) = \frac{1}{3}I^{v,p}(\theta) + I^{h,p}(\theta) + \frac{2}{3}I^{h,s}(\theta). \quad [6]$$

The extraction efficiency within the critical angle is

$$\eta_{ext} = \frac{\int_0^{\theta_c} I(\theta)\sin(\theta)d\theta}{\int_0^{2\pi} I(\theta)\sin(\theta)d\theta}, \quad [7]$$

which can be evaluated by a discrete summation.[8] The angular dependence of the reflection, absorption, and transmission of each layer was calculated by the transfer matrix method. To illustrate the resonant cavity effect, two systems were modeled as displayed in Figure 1. Specifically, a GaN micro-cavity is bound on either side by a metallic mirror and a 7x GaN / AlN Bragg mirror.[12] Figure 1(a) presents the straight-forward design of a semi-transparent Ni / Au (2 nm / 6 nm) top contact and a DBR on Si bottom mirror ($R_2$=94%). In contrast, Figure 1(b) presents the same structure but in a flip-chip design with a thick Ag layer ($R_2$=97%) as the bottom mirror and a 7x DBR as the top mirror. A wet etch can selectively remove the Si substrate to allow transmission out of the DBR structure. The reflectance of the 7x Bragg mirror without the underlying Si substrate decreases to approximately 65%.[13]

Theoretical calculations predict that a resonant cavity created with this DBR structure and a top-side external metal mirror would yield a greater than 4x improvement (Figure 1(a)) and a within a flip chip configuration would yield a greater than 8x increase in the extraction efficiency of the LED (Figure 1(b)). The flip chip design is more complicated, but the higher reflectance from the bottom metallic mirror and the lower absorbance of the top DBR mirror provides a better optical design. Still, the major incentive for GaN-based LEDs on Si is the cost advantage gained with processing larger substrates; however, an ultra-high brightness LED product motivates a costly flip-chip design. Figure 1(a) does show that resonant cavity enhancement can be achieved despite the lower reflectance ($R_1$=40%) in the GaN / Ni / Au / air structure.

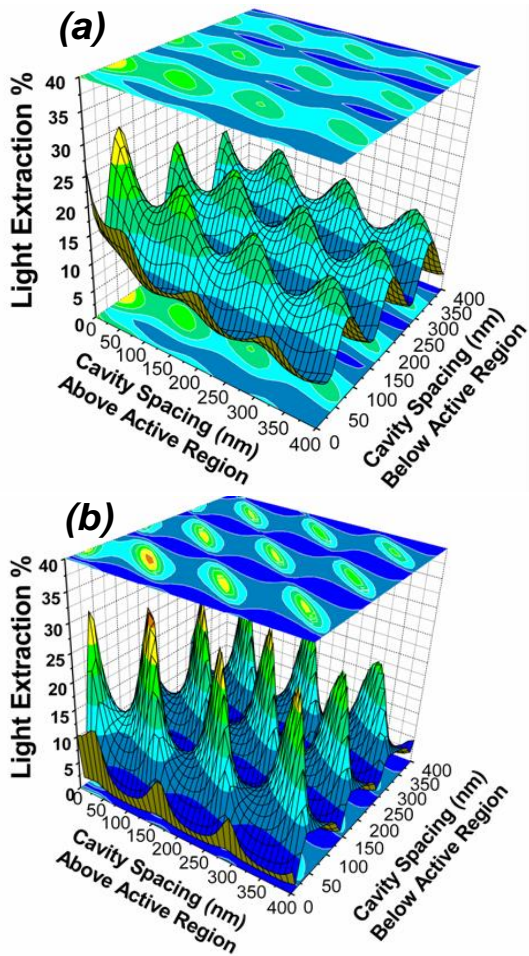

Figure 1. Extraction efficiency predicted for a dipole source placed in a GaN micro-cavity surrounded by (a) a top semi-reflective metallic contact and a bottom 7x GaN / AlN Bragg mirror on a Si substrate and (b) a flip-chip design with a top 7x GaN / AlN Bragg mirror and a thick Ag bottom mirror.

Several samples consisting of a single quantum well (SQW) UV-LED on a 7x GaN / AlN DBR on a Si substrate were grown in a MOCVD system. The LED structure consisting of a 100-nm GaN:Si, a 20-nm / 10-nm / 20-nm AlGaN:Si / GaN-undoped / AlGaN:Mg SQW, and a 125-nm GaN:Mg layer were constant in all samples studied, while the thickness of the GaN and AlN layers in the DBR were adjusted to set the peak reflectance to various wavelengths in the UV and visible light regime. A top Ni or Ni / Au contact provided contact to the p-GaN, while the silver paste was applied following an HF etch and SC1 clean of the conductive $n^+$ Si substrate to remove the native oxide to provide a reliable backside contact.

The electroluminescence (EL) spectrum presented in Figure 2(a) displays the UV, blue (425 nm) and yellow (550 nm) spectral peaks. The band-edge transition, located near 365 nm, dominates although this peak also includes lower-energy contributions from the zero-phonon free-electron to Mg acceptor (e0-A) transition at approximately 380 nm and donor-acceptor pair transitions at 395 and 410 nm.[14] A blue transition was also observed in the GaN:Mg layer that occurs near 430 nm via a deep donor, e.g., nitrogen vacancy, to shallow acceptor pair transition. Additionally, a weak transition is observable at 550 nm from the well-known GaN mid-gap yellow luminescence that is due to a shallow donor to deep acceptor transition,[15] which is dependent upon the presence of a nitrogen or gallium vacancy.[16]

Figure 2(b) displays an EL spectrum acquired from an LED coupled to a DBR with a reflection band-width centered at 385 nm, near the dominant band edge emission region. The UV emission was improved by 2.5 times by the resonant cavity when compared to an identical SQW-LED reference device grown on a standard

GaN on Si structure. The spectral-dependence of DBR and resonant-cavity structures provides a simple method to suppress off-resonant emission levels to allow for improved spectral purity. In Figure 2(b), the donor to acceptor blue (425 nm) and the yellow (550 nm) mid-gap defect-based transition were both suppressed. As expected in Figure 2(a), EL measurements taken from an identical SQW-LED device grown on mismatched DBR, i.e., DBR structures with a reflection band-width in the blue-green portion of the visible spectrum, displayed a strong enhancement in this regime.

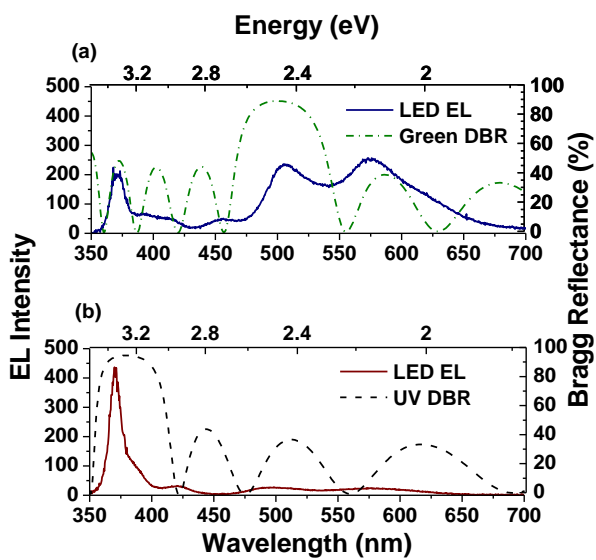

Figure 2. Electroluminescence spectra (solid line) from identical SQW-UV LEDs at a current injection level of 50 mA with corresponding DBR reflectance (dashed line) centered in the (a) UV and (b) blue-green

The emission from the LEDs was clearly modified by the resonant cavity. Still, several additional experimental factors may also influence the model presented above. The calculations only accounted for light extracted from within the light cone. In commercial LEDs, a significant portion of the emitted light is guided internally within the structure and eventually transmitted and collected at the edge of the die. Also, guided light may be recycled, i.e., reabsorbed / reemitted, in the active region of the LED. This three-dimensional model is under development. The angle dependent reflectance from the GaN / AlN structure as well as a 5x and 9x DBR on a Si substrate is displayed in Figure 3. The DBR reflection band-width clearly overlaps the light extraction cone for s- and p-polarization. It is more essential to understand the nature of lateral light propagation (at approximately 90°) particularly when accounting for light transmitted from the edge of the die.

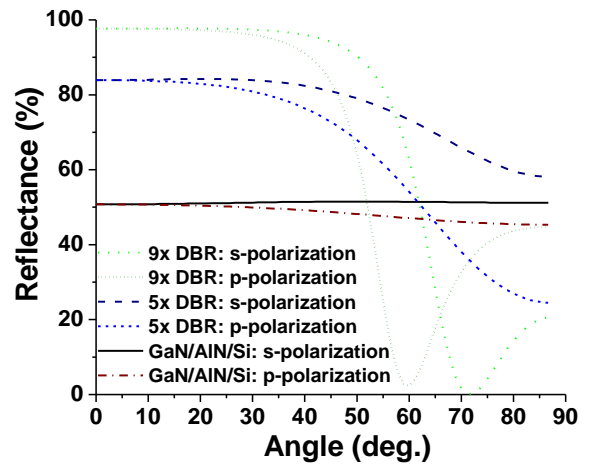

Figure 3. Angle dependent reflectance from GaN / AlN structures on a Si substrate. For reference, the light extraction cone at the GaN / air interface extends to 23.98°.

Only recently was it reported by our group that a crack-free III-nitride DBR can be grown on Si with a reflectivity over 97% by exploiting the large index of refraction contrast with the substrate.[17] The low number of periods in this structure inherently produced a reflection band-width that extends over a large wavelength range.[11]

The standard production process for GaN requires growing thick layers on sapphire or SiC substrates.[18,19] An approximately 40 period GaN/AlN DBR is required to achieve high reflectance on this structure.[20] In addition to the cracking dilemma associated with this structure, the reflection band-width is extremely narrow for a DBR with such a large number of periods. Thus, achieving enhanced emission from an optical cavity based on this structure is highly dependent on optimal placement of the active layer and achieving a light source with a spectral width near the theoretical limit for GaN.[18] In contrast, this letter reveals that extracted emission from GaN-based LEDs can be significantly enhanced by exploiting this elegant GaN/AlN Bragg reflector on Si design.

*Experimental*

This letter reports on the metalorganic chemical vapor deposition (MOCVD) of RCLED devices based on an AlN / GaN superlattice structures grown directly on Si(111) substrates. The alternating sequence functions optically as a high-reflectance DBR and structurally as a strain compensating superlattice. The growth was carried out in a modified vertical-impinging-flow chemical-vapor-deposition reactor. Two-inch Si wafers were cleaned via a modified Radio Corporation of America process followed by an in situ $H_2$ bake. An Al seed layer was deposited prior to the onset of $NH_3$ flow to protect the Si surface from nitridization. The AlN/GaN superlattice was deposited at 1050°C and 50 Torr. The reflection band-width for the DBR structure can be tuned to any point in the near-UV or visible by adjusted the thicknesses of the AlN and GaN layers in the DBR. For example, the DBR with a reflection band-width centered at 385 nm (UV) required alternating 45.6-nm AlN and 43.0-nm GaN layers. Following growth of the DBR layers, an identical GaN SQW LED device was deposited at 1020°C at 150 Torr directly on the various DBR structures, as well as on a control GaN on Si structure. The n- and p-type doping was accomplished with disilane and $Cp_2Mg$, respectively. Topside metal contacts were deposited and patterned on the p-type GaN layer along with blanket silver contact deposition on the backside of the Si wafer.

Optical characterization of the various LEDs was performed on multiple diodes with varying DBR thicknesses, in an effort to quantify the overall spectral output and its dependence on the DBR properties. Circular nickel and gold metal contacts were deposited across the top surface of the device and silver paste was used to make contact to the backside, following an HF etch and SC1 (1:1:5 $NH_4OH:H_2O_2:H_2O$) clean to remove any native oxide that had formed to ensure the best possible contact was made. A constant current between 2.5 and 200 mA was injected into the LED via the top metal contact using a Keithley 2400 Source Measurement Unit. EL spectra were collected using a Mitutoyo microscope and 100X (0.7 NA) objective and the spectral output were directed via an optical fiber into an Ocean Optics USB2000 spectrometer with an effective spectral range of 200 to 850 nm. Non-linearities of the spectrometer were corrected for via comparison of a measurement of the output of a tungsten lamp with a calculated 3100K color temperature spectral plot, to ensure that accurate intensity comparisons were obtained. The *ex situ* reflectance was measured at normal incidence using a tungsten lamp as a source and the reflected beam was dispersed through an Ocean Optics USB2000 spectrometer with a 50-μm slit.